\documentclass[twocolumn,aps,prl,lengthcheck,showpacs]{revtex4}
\usepackage[dvips]{graphicx}
\usepackage{latexsym}
\usepackage{epsfig}
\usepackage{bm}
\usepackage{times,psfrag,subfigure}   
\usepackage{color}     
\begin{document}

\title{Non-isomorphic nucleation pathways arising from morphological transitions
of liquid channels}
\author{H. Kusumaatmaja, R. Lipowsky, C. Jin, R. -C. Mutihac and H. Riegler}
\affiliation{Max Planck Institute of Colloids and Interfaces, Science Park Golm, 14424 Potsdam, Germany}
\date{\today}

\begin{abstract}

Motivated by unexpected morphologies of the emerging liquid phase (channels, bulges, droplets) at the edge of thin, melting alkane terraces, we propose a new heterogeneous nucleation pathway. The competition between bulk and interfacial energies and the boundary conditions determine the growth and shape of the liquid phase at the edge of the solid alkane terraces. Calculations and experiments reveal a ``pre-critical'' shape transition (channel-to-bulges) of the liquid before reaching its critical volume along a putative shape-conserving path. Bulk liquid emerges from the new shape, and depending on the degree of supersaturation, the new pathway may have two, one, or zero energy barriers. The findings are broadly relevant for many heterogeneous nucleation processes because the novel pathway is induced by common, widespread surface topologies (scratches, steps, etc.).
\end{abstract}
\maketitle


{\it{Introduction.}}
Phase transitions are influenced by the size or dimensionality of the system, because of contributions from interfacial free energies  \cite{Christenson,Dash}. While this has long been known, many details are still poorly understood. Here we analyze the melting process of such a confined system, the melting of terraced solids. Our findings show for the first time the relevance of morphological instabilities for nucleation processes, in particular how pre-critical morphological transitions can lower the nucleation barriers.

Recently, it was shown that the melting of monolayer terraces of long chain alkanes at planar silica/air interfaces \cite{Lazar} leads to moving alkane drops. The drop movement was explained by the wetting properties of the system. However, the emergence of the drops before they start moving has never been investigated. In this paper we present new experimental data and analyze for the first time theoretically the onset of the drop formation.

The emergence of the new phase is associated with a nucleation process. Classical nucleation theory \cite{nucleation} shows that for alkanes the critical (heterogeneous) nucleus size is of order 100 nm \cite{nucleationdata}. This is much smaller than the observed drop sizes (microns, Fig. 1). The question arises what mechanism leads to the unexpectedly large incipient droplets. We show here that the transition to the macroscopically emerging and growing new phase is critically coupled to a morphological transition or capillary instability from uniform to bulged liquid channels at the terrace edges. The instability provides an energetically favorable pathway for the phase transition. Our theory further predicts that this instability can only occur when the terrace edge length is sufficiently large compared to the edge height. This conjecture is confirmed by experiments on the melting of alkane multilayer island terraces of varying thicknesses.

Apart from addressing the relevance of morphological instabilities for nucleation processes, our work shows that surface ledges on  the nanometer scale can induce droplets that are orders of magnitude larger, which differs from previous observations of channel-to-droplet instabilities on surface stripes \cite{Gau}. Long-scale instabilities have been observed in thin layers, see e.g.  \cite{Craster,wetting}, but they typically involved external forces in contrast to the systems studied here.

Our system has rather general properties: (i) melting starts at the edge, (ii) a low contact angle of the melt with the surface, and (iii) the contact line is pinned at the terrace edge. It describes widespread situations such as the melting of flat (2-dimensional) solid domains or condensation/adsorption processes on surfaces with terraces, steps, etc. It is also technologically relevant \cite{wetting,microfluidics,Seemann,Courbin}.

{\it{Experimental preparation}}
\begin{figure}
\centering
\includegraphics[scale=0.975,angle=0]{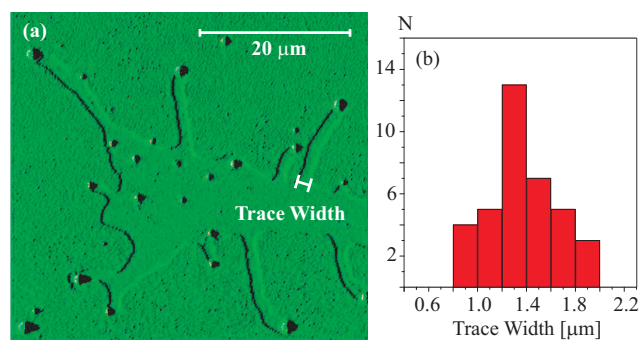}
\caption{(color online) (a) Moving alkane droplets appear above the monolayer melting temperature. (b) Histogram of the droplet size. The size of the droplet is equivalent to the trace width.}
\label{sizeaverage}
\end{figure}
We prepared planar silica surfaces, which were partially covered with large (tens of microns in lateral dimension), contingent areas or terraces of uniform thickness that are formed by mono- or multilayers of solid long chain n-alkanes $\mathrm{C_{36}H_{74}}$. The thickness of a solid alkane layer is approximately identical to the molecular all-trans length ($\simeq 4.8$ nm; the molecules are oriented upright \cite{Merkl,Holzwarth}). For more details on the substrate preparation see \cite{Kohler} and \cite{SI}. 

{\it{Experimental observation on monolayer melting.}}
If the temperature is gradually raised above the melting point of the alkane monolayer $T_m$ \cite{noteexp}, eventually alkane drops appear at the terrace edges and start to move through the solid monolayer area \cite{KohlerR}. The drops have a small contact angle, $\theta \leq 15^\circ$. We focus here on the early stages of the drop formation. We observe that the drops are always already of micrometer size as soon as they appear at the edge (see supporting movie). This is unexpected because the step height is only nanometers. The drop size is measured via atomic force microscopy (AFM) by cooling samples with moving drops very quickly, thus freezing the configuration (Fig. \ref{sizeaverage}). The typical width of the orifices of the paths of the drops is $\simeq 1.3 \mu$m (histogram, Fig. \ref{sizeaverage}), about 300 times larger than the height of the terraces!

{\it{Theoretical model.}}
\begin{figure}
\centering
\includegraphics[scale=0.35,angle=0]{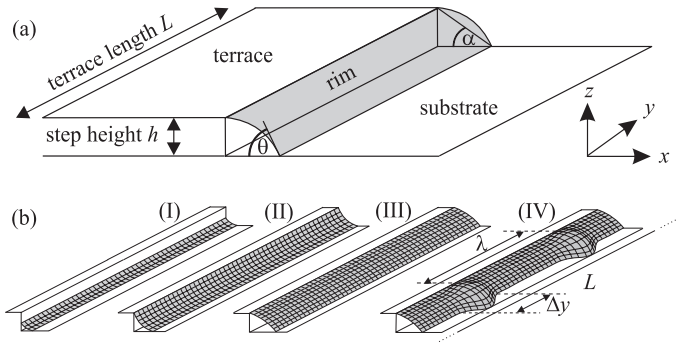}
\caption{(a) Schematic diagram of the geometry: $\theta$ is the contact angle and $\alpha$ is the angle between the diagonal and horizontal lines. (b) Possible morphologies for a given contact angle and increasing liquid volume: (I-II) a uniform channel with negative curvature, (III) a uniform channel with positive curvature, and (IV) a bulged channel. In (IV), the liquid bulge is not to scale. It is typically much larger.}
\label{geometry}
\end{figure}
Fig. \ref{geometry}(a) shows the geometry of the experimental configuration of the alkane melting from the terrace edge with a step of height $h$ and length $L \gg h$ between the smooth and homogeneous terrace and substrate surfaces. The liquid wets the substrate, rim, and terrace surfaces partially with contact angle $\theta$. The angle $\alpha$ in Fig. \ref{geometry}(a) is a measure of the width of the liquid channel. 

The competition between surface and bulk energies determines how much liquid is present at temperature $T$ and which morphology will be formed. The increase in the volume of liquid alkane corresponds to the decrease in the volume of the solid alkane layers (the densities of liquid and solid alkane are approximately equal). Due to interfacial energy contributions and in agreement with experimental observations, alkane melt will first appear at the terrace ledge. Depending on the liquid volume $V$, for a given (low) contact angle $\theta$, there are four distinct morphologies (Fig. \ref{geometry}(b)). (I) A channel with negative curvature at the liquid/vapour surface and an unpinned contact line. (II) Like (I), but with the (left segment) contact line pinned at the terrace edge. (III) A channel with positive curvature, pinned at the terrace edge. (IV) A channel with bulges (localised drops), still pinned at the terrace edge, with a periodic configuration in the $y$-direction (bulge-to-bulge distance $\lambda > \Delta{}y$, $L = n \lambda$; $\Delta{}y$ is the size of the droplet and $n$ is an integer). The contact line depins when the angle of the liquid with the terrace surface exceeds the equilibrium contact angle.

The interfacial energy of the system can be written as 
\begin{equation}
E_I = \Sigma_{LG} A_{LG} - \Sigma_{LG}\cos{\theta}A_{SL}, \label{free}
\end{equation}
$\Sigma$ is the surface tension, $A$ the area. The subscripts $L$, $G$ and $S$ denote liquid, gas, and solid. The relation
\begin{equation}
\Sigma_{LG}\cos{\theta} = \Sigma_{SG}-\Sigma_{SL}
\end{equation}
defines the contact angle $\theta$. Molecular effects are neglected, since as we shall see later, the size of the resulting droplet is of order of microns \cite{Santer}. 

The interfacial areas of morphologies II and III, and the liquid volume are calculated analytically:
\begin{eqnarray}
& A_{LG} = \frac{hL}{\sin{\alpha}} \frac{(\theta-\alpha)}{\sin{(\theta-\alpha)}} ; \;\; A_{SL} = \frac{hL}{\tan{\alpha}} , \label{Area} \\
&V/h^2L = \frac{1}{2\tan{\alpha}} + \frac{\left((\theta-\alpha) - \sin{(\theta-\alpha)}\cos{(\theta-\alpha)}\right)}{4\sin^2{\alpha}\sin^2{(\theta-\alpha)}}. \label{volume}
\end{eqnarray}
Substituting Eq. (\ref{Area}) into (\ref{free}) yields the interfacial energy. To first approximation, we shall assume the terrace length, and thus the surface energy of the vertical ledge, to be constant. For morphology IV, we use a numerical method (Surface Evolver \cite{Brakke}). It minimises the surface energy of a given volume of liquid with a prescribed geometry. Surface tensions and contact angles are inputs.

Bulk energy contributions,  $\Delta\mu V$, result from melting the solid, with $\Delta\mu = \mu_l(T,P)-\mu_s(T,P)$, and $\mu_l(T,P)$ and $\mu_s(T,P)$ as chemical potentials per unit volume in the liquid and solid phases, respectively. We estimate \cite{Dash}
\begin{equation}
\Delta\mu \simeq -q_m(T-T_m)/T_m , \label{bulkeq}
\end{equation}
with $q_m$ the latent heat of melting per unit volume. Positive and negative $\Delta \mu$ correspond to supersaturation of the liquid and solid, respectively.

In the following we shall use normalized variables $\bar{E}=E/\Sigma_{LG}hL$, $\bar{V}=V/h^2L$, $\bar{P_L}=P_Lh/\Sigma_{LG}$, $\Delta\bar{\mu}=\Delta\mu h/\Sigma_{LG}$, and $\bar{\lambda} = \lambda/h$.

{\it{Coexistence of channel and bulged morphologies.}}
Simulations reveal that for a system as shown in Fig. \ref{geometry}(a), besides uniform channels, other isocurvature shapes are possible (i.e., shapes with uniform pressure within the liquid). These shapes are characterized by bulges of equal size $\Delta{}\bar{y}$ that are separated by identical channel sections with distances $\bar{\lambda} > \Delta{}\bar{y}$, see morphology IV in Fig. \ref{geometry}.

\begin{figure}
\centering
\includegraphics[scale=0.9,angle=0]{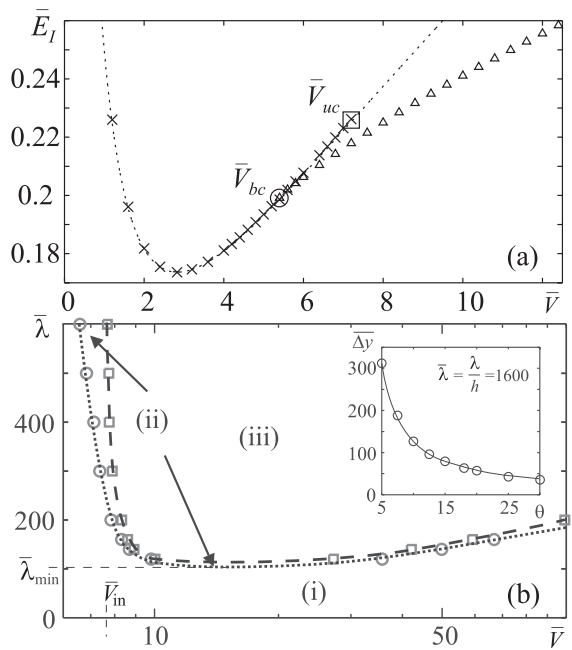}
\caption{(a) Interfacial energy as function of volume for uniform and bulged morphologies (dashed line = analytical solution; crosses (uniform) and triangles (bulged) = numerical results). Here we use $\bar{\lambda} = 1600$ and $\theta=10^\circ$. $\bar{V}_{uc}$ and $\bar{V}_{bc}$ are instability points for the uniform and bulged morphologies respectively. (b) Morphology diagram showing regions where only uniform channels (i), bulges (iii) or both morphologies (ii) can exist. The inset in (b) shows the bulge/drop dimension $\Delta\bar{y}=\Delta y/h$ as a function of the contact angle $\theta$.}
\label{EInterface}
\end{figure}

Fig. \ref{EInterface}(a) shows the normalized interfacial energy of the system $\bar{E}_I$ as a function of volume $\bar{V}$ for both, uniform and bulged morphologies, respectively ($\theta = 10^\circ$ and $\bar{\lambda} = 1600$, representing our experimental observations). At small volume, the system is in the uniform channel morphology. The energy minimum describes a metastable uniform channel with zero curvature (zero Laplace pressure). With increasing volume, at the minimum, the channel curvature changes  from negative to positive.

At a volume $\bar{V}_{uc}$ the uniform morphology becomes absolutely unstable and instantaneously transforms into a bulged morphology (morphology IV, triangles). This transition at $\bar{V}_{uc}$ is linked to a {\it{jump}} in energy. There is also a lower instability volume $\bar{V}_{bc}$, where starting in the bulged shape with $\bar{V} > \bar{V}_{bc}$, upon reducing the volume, the system will retain its shape until it transforms into a uniform channel at $\bar{V}_{bc}$. Between $\bar{V}_{uc}$ and $\bar{V}_{bc}$ the system behaves hysteretically. 

Fig. \ref{EInterface}(b) summarizes the results of calculations as depicted in Fig. \ref{EInterface}(a) for different values of $\bar{\lambda}$ in a morphology diagram. It shows regions where only uniform channels are possible (i), both bulges and uniform channels can exist (ii), and only bulges are possible (iii). The boundary between regions (ii) and (iii) also agree well with an analytical linear instability analysis \cite{SI}, which shows that the lowest volume instability occurs at $\alpha = \theta/2$ i.e., at $\bar{V}_{\mathrm{in}} = 1/2\tan(\theta/2) + (\theta/2 - \sin(\theta)/2)/4\sin^4(\theta/2)$, the volume corresponding to the maximum Laplace pressure $\bar{P}_L$ of the bulged channel, as we shall see below.

\begin{figure}
\centering
\includegraphics[scale=1.1,angle=0]{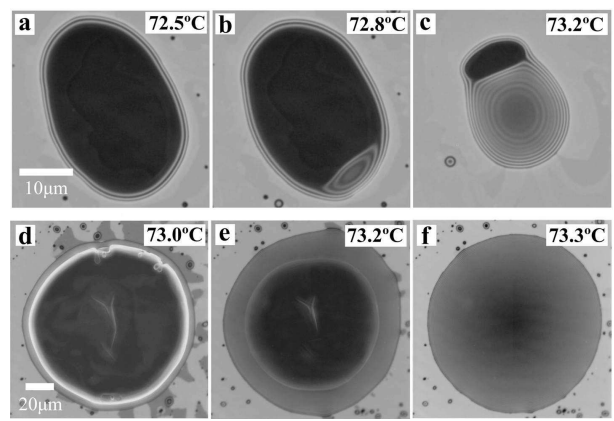}
\caption{Melting behaviour of solid alkane terraces with heights of $\simeq 500$ nm (upper row, a-c) and $\simeq 4 \mu$m (lower row, d-f). The 500 nm terrace first grows a liquid channel (a) that eventually destabilizes into a growing bulge (b, c). The $\simeq 4 \mu$m terrace also grows a liquid channel (d). It remains a channel (e) that eventually converges into a drop (f) without prior bulge formation.}
\label{expfig}
\end{figure}

Quite remarkably Fig. \ref{EInterface}(b) reveals an absolute lower limit for $\bar{\lambda}$. For a bulged channel, the quasi-cylindrical sleeves of the channel must have the same mean curvature as the quasi-spherical bulge. In order to obtain a sphere with the same mean curvature as a cylinder, the curvature radius of the sphere must be twice the curvature radius of the cylinder. Thus, we need a certain minimal size of the bulge, and a certain minimal amount of volume which we can extract from the quasi-cylindrical sleeves to form the bulge. The existence of a $\bar{\lambda}_{\mathrm{min}}$ also means melting will proceed very differently depending on the available normalized length $\bar{L} = L/h$. Experimentally this means, the melting behavior of alkane terraces should depend on their perimeter $L$ respectively height $h$. This is indeed what we observed Fig. \ref{expfig} (see also movies in \cite{SI}). In the first row, $L/h \simeq 200 > \bar{\lambda}_{\mathrm{min}}$, and bulged morphology appears. On the other hand, $L/h \simeq 100 < \bar{\lambda}_{\mathrm{min}}$ in the second row and the system melts in a channel shape. We note that energy minimization alone favors the formation of just a single bulge (i.e. $\lambda = L$). However, defects and dynamical effects may favor the formation of several droplets. Detailed descriptions of Fig. \ref{expfig} are given in the supporting information.

The inset in Fig. \ref{EInterface}(b) shows the corresponding bulge size $\Delta{}\bar{y}$ for a fixed $\bar{\lambda}=1600$ (other $\bar{\lambda}$ give similar results) and various contact angles when the bulge morphologies first become the global minima. $\Delta{}\bar{y}$ represents the minimum moving drop dimensions because if the system forms bulged shapes as the temperature increases (as shown below), these bulges will grow in size and eventually become drops. The calculated $\Delta{}y\simeq 600$ nm for our experimental conditions ($h \simeq 5$ nm, $\theta \simeq 10^\circ$) agrees quite well with the experimentally observed minimum drop size.

{\it{Nucleation barrier.}} For the nucleation and melting pathway, we need the total energy of the system $\bar{E} = \bar{E}_I + \Delta {\bar{\mu}} \bar{V}$. Stationary shapes correspond to $d\bar{E}/d\bar{V}=0$, with $d\bar{E}_I/d\bar{V} = \bar{P}_L$, this means
\begin{equation}
\Delta\bar{\mu} + \bar{P}_L = 0. \label{eqcon}
\end{equation}

\begin{figure}
\centering
\includegraphics[scale=0.85,angle=0]{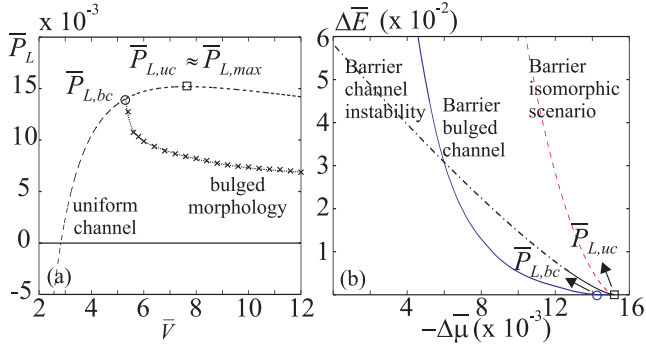}
\caption{(a) Laplace pressure as a function of volume for uniform (dashed line, analytical) and bulged (crosses, Surface Evolver) morphologies. (b) Nucleation energy barriers as a function of supersaturation. $P_{L,\mathrm{uc}}$ and $P_{L,\mathrm{bc}}$ are the Laplace pressure at the instability points for the uniform and bulged channels.}  
\label{Laplace}
\end{figure}

For nucleation on a uniform substrate surface, the nucleation pathway is provided by a spherical cap that is isomorphically upscaled. Now, let us consider, for the moment, a corresponding nucleation scenario as obtained by isomorphic upscaling of the uniform channel, a scenario that ignores the morphological instability as illustrated in  Fig. 3(a). For the uniform channel, $\bar{P}_L =  2\sin{\alpha}\sin{(\theta-\alpha)}$.  $\bar{P}_L$ is plotted in Fig. \ref{Laplace}(a) (dashed line). The Laplace pressure has a maximum, $\bar{P}_{L,{max}}$, at $\alpha = \theta/2$. This implies (Eq. (\ref{eqcon})) a critical temperature, $T_{\mathrm{in}}$, above which there is no energy barrier to complete melting. $T_{\mathrm{in}}$ is given by
\begin{equation}
T_{\mathrm{in}} = T_m \left[ 1 + \frac{2\Sigma_{LG}}{hq_m}\sin^2{\frac{\theta}{2}} \right]. \label{criT}
\end{equation}
We estimate $T_{\mathrm{in}}-T_m \sim 0.2^\circ$ with typical values for $C_{36}H_{74}$ \cite{nucleationdata}.
Between $0 < \bar{P}_L < \bar{P}_{L,{max}}$, there are two channel volumes $\bar{V}$ with the same pressure. Hence there are two solutions to Eq. (\ref{eqcon}) for a given supersaturation $\Delta\bar{\mu}$. The solution with the lower $\bar{V}$ corresponds to a local minimum. The larger $\bar{V}$ corresponds to a local maximum. Regarding the nucleation barrier, the larger $\bar{V}$ signifies the peak of the energy barrier for complete melting. The energy barrier for this classical isomorphic nucleation scenario is shown as dashed line in Fig. \ref{Laplace}(b). 

The isomorphic nucleation scenario just discussed only applies if the uniform channel does not undergo a morphological transition. In particular, in the situation considered here, a channel with $\bar{\lambda} > \bar{\lambda}_{\mathrm{min}}$  will become unstable due to the morphological (bulging) instability discussed in the preceding section \textit{\textbf{before}} it reaches its critical nucleation volume as a channel.  
A  \textit{\textbf{"precritical"}} morphological transition opens a new and lower energy nucleation pathway. The Laplace pressure of the bulged shapes is plotted in Fig. \ref{Laplace}(a) for $\bar{\lambda} = 1600$ (crossed symbols). 
There are three different regimes of supersaturation. (i) For $-\Delta\bar{\mu} < \bar{P}_{L,{bc}}$, the nucleation free energy landscape has two barriers. The first barrier corresponds to  the upper instability point, at which the uniform channel decays into the bulged channel. This barrier, shown as a dashed-dotted line in Fig. \ref{Laplace}(b), is not a smooth maximum but instead a sawtooth-shaped discontinuity. The second barrier (solid line) arises from the bulged channel solution. Using the same argument as before, the peak of the barrier corresponds to the solution of Eq. (\ref{eqcon}) with $P_L$ now the Laplace pressure of the bulged morphology. For small positive $-\Delta\bar{\mu}$, the second barrier is much larger than the first one and dominates the nucleation process. As we increase $-\Delta\bar{\mu}$ towards $\bar{P}_{L,{bc}}$, the second barrier of the nucleation free energy is reduced and eventually disappears. With increasing $\lambda$, the value of $\bar{P}_{L,{bc}}$ decreases. It is, however, very difficult to numerically determine its asymptotic value. (ii) For $\bar{P}_{L,{bc}} < -\Delta\bar{\mu} < \bar{P}_{L,{uc}}$, there is only a single barrier arising from the upper instability point of the uniform channel. This barrier disappears at $-\Delta\bar{\mu}=\bar{P}_{L,{uc}}$. It is worth noting that from our linear instability analysis \cite{SI}, we find in the limit of large $\bar{\lambda}$, $\bar{P}_{L,{uc}} \simeq \bar{P}_{L,{max}}$. (iii) Finally, the nucleation free energy does not exhibit any barrier for $-\Delta\bar{\mu} > \bar{P}_{L,{uc}}$. 

{\it{Discussion and Conclusion.}}
We explain the formation of liquid channels at the edges of alkane terraces, their morphological transformation into bulges and the observation of unexpectedly large incipient alkane drops. 
We present a general nucleation and growth scenario for the melting of solid terraces if the liquid melt does not completely wet the surfaces. Promoted by interfacial energy contributions, melting begins already below the bulk melting temperature $T_m$. The liquid forms a uniform channel whose surface curvatures changes from negative to positive when the temperature exceeds $T_m$. Above  $T_m$ the liquid channel is meta-stable because the interfacial energy suppresses complete melting. If the uniform channel grew isomorphically, it would reach a certain critical volume corresponding to a maximum of the total energy. Then the continuously growing bulk phase emerges. We show here for the first time in experiment and theory that under certain conditions (sufficiently long terrace edges) the growing channel transforms into a bulged morphology \textit{\textbf{before}} it reaches its critical nucleation volume as a channel. This "pre-critical" morphological transition opens a new nucleation pathway for the system. Its nucleation barrier is significantly lower than that obtained by isomorphic upscaling of the uniform channel geometry. This is an important result since the probabilities to overcome the nucleation barrier typically follow exponential rules $\propto e^{-\Delta{}E/kT}$.
This instability aspect distinguishes our work from previous ``multistep'' nucleation mechanisms. For examples, nucleation pathways with two energy barriers but no shape instabilities have been reported \cite{Valencia}; during the crystallization of silicon, it was shown that the critical (crystalline) nucleus is preceded by the formation of high-density liquid polymorph \cite{silicon}. Our investigations further reveal interesting relations between the bulging morphology and the terrace geometry. 

Our results are quite general and relevant beyond the melting of alkane terraces. The terrace edge is a paradigm for ubiquitous  surface "roughness" topologies, natural or artificial (scratches). The described growth scenario of emerging bulk phases upon melting analogously holds for many condensation, adsorption, precipitation or deposition processes. It suggests, for instance, that a supersaturated vapor preferentially forms macroscopic liquid drops at a long edge before this happens at a short edge of the same height.

{\it{Acknowledgements}} We thank H. M\"ohwald and P. Lazar for useful discussions. R.C.M. and C.J. acknowledge support from IMPRS on Biomimetic Systems.



\begin{thebibliography}{99}
\bibitem{noteexp} Long chain alkanes show surface freezing. The melting point of the first monolayer at the substrate surface is a few degrees higher than the macroscopic bulk melting point \cite{Merkl}. For our purpose, it simply means the suitable reference melting temperature is the monolayer melting temperature $T_m$. The described melting scenario is rather general and can be observed for various alkane steps \cite{Lazar}.
\end{thebibliography}
\end{document}